\documentclass[12pt,preprint]{aastex}
\usepackage{psfig,lscape}
\def\lap{\lower.5ex\hbox{$\; \buildrel < \over \sim \;$}}
\def\gap{\lower.5ex\hbox{$\; \buildrel > \over \sim \;$}}

\def\ergcm2s{${\rm erg\ cm^{-2}\ s^{-1}}$}

\def\ergscm2s{${\rm erg\ cm^{-2}\  s^{-1}}$}

\def\cm-2{${\rm cm^{-2}}$}
\def\ergs{${\rm erg\ s^{-1}}$}

\begin{document}

\title{An X-ray Transient and Optical Counterpart in the M31 Bulge}

\author{Benjamin F. Williams\altaffilmark{1}, Michael
R. Garcia\altaffilmark{1}, Jeffrey E. McClintock\altaffilmark{1},
Albert K. H. Kong\altaffilmark{1}, Frank A. Primini\altaffilmark{1},
and Stephen S. Murray\altaffilmark{1}}
\altaffiltext{1}{Harvard-Smithsonian Center for Astrophysics, 60
Garden Street, Cambridge, MA 02138; williams@head.cfa.harvard.edu;
garcia@head.cfa.harvard.edu; jem@head.cfa.harvard.edu;
akong@head.cfa.harvard.edu; fap@head.cfa.harvard.edu;
ssm@head.cfa.harvard.edu }


\begin{abstract}

We have obtained snapshot images of a transient X-ray source in M31
from {\it Chandra} ACIS-I and the {\it Hubble Space Telescope (HST)}
Advanced Camera for Surveys (ACS).  The {\it Chandra} position of the
X-ray nova was R.A.=00:42:56.038 $\pm$0.08$''$, Dec.=+41:12:18.50
$\pm$0.07$''$.  The transient was active for at least 6 months.
Previous observations set an upper limit before the X-ray outburst,
demonstrating variability by a factor of $>$100 and confirming the
transient nature of the source.  For the first 6 months after the
initial detection, the X-ray luminosity was $\sim$6$\times$10$^{37}$
erg s$^{-1}$; it then decayed to $<5\times$10$^{36}$ erg s$^{-1}$ over
the following 2 months.  An {\it HST} observation 29 days after the
initial X-ray detection revealed a source at R.A.=00:42:56.042,
Dec.=+41:12:18.45 that was $B=24.52\pm0.07$.  This optical source
faded to $B=24.95\pm0.08$ in 9 months. The {\it HST} identification of
an optical source at the same position as the X-ray source, fading in
concert with the X-ray source, indicates that this optical source is
the counterpart of the X-ray transient.  The lack of high-mass stars
in the region suggests this source is a low-mass X-ray binary, and the
X-ray and optical luminosities provide a rough orbital period estimate
of 8$^{+12}_{-5}$ days for the system.

\end{abstract}

\section{Introduction}

Most Galactic soft X-ray transient (SXT) sources in low mass X-ray
binaries show compelling dynamical evidence that they contain a
stellar-mass black hole \citep{mcclintock2004}.  The variability on
millisecond timescales and short-term X-ray luminosities reaching
$>$10$^{38}$ erg s$^{-1}$ provide further support that these are the
some of the most likely black hole candidates known
\citep{charles1998}.  Such objects are therefore of intense interest
as sites for more detailed studies of general relativity.  Finding
such sources in the Galaxy requires all-sky monitoring and difficult
distance estimates.  Searches for bright transient X-ray sources in
nearby galaxies, such as M31, therefore provide a very efficient
method for expanding the known sample of these fascinating sources.

Since the M31 bulge is at a known distance, has low extinction, and
can be surveyed in a single {\it Chandra} observation, it presents an
excellent laboratory for searching for transient X-ray sources.  Such
transient sources appear about once each month in the M31 bulge alone
\citep{williams2004}, and M31 surveys have already discovered nearly
fifty transient X-ray sources
(\citealp{trudolyubov2001,kong2002,distefano2004,williams2004,williams2005a}).

Furthermore, the exquisite angular resolution of the {\it Hubble Space
Telescope (HST)}, allows the individual stars in the M31 bulge to be
resolved.  Combining this resolution with the positional accuracy of
{\it Chandra} allows one to search for optical counterparts for
transient X-ray sources in the M31 bulge.  Presently, optical
counterpart candidates have been identified from {\it HST} for three
transient X-ray sources in M31
\citep{williams2004,williams2005a}. Herein, we report the discovery of
a new transient X-ray source in the M31 bulge found by our {\it
Chandra/HST} monitoring campaign.  Exceptional positional accuracy and
image alignments for the nearly contemporaneous {\it Chandra} and {\it
HST} data sets provide a reliable identification of the optical
counterpart.

\section{X-ray Data}

We obtained {\it Chandra} ACIS-I images of the M31 bulge on
26-Nov-2003, 27-Dec-2003, 31-Jan-2004, 23-May-2004, and 17-July-2004.
The observations were performed in ``alternating exposure readout'',
so that every 6th frame had 0.6 seconds of exposure instead of the
canonical 3.2 seconds.  This mode lowers the effective exposure time
by $\sim$20\%, but it provides a second low exposure image in which
bright sources are not piled up.  The details of these observations,
including target coordinates, roll angle of the telescope, and
exposure time, are provided in Table~\ref{xobs}.

These X-ray observations were all reduced in an identical manner using
the X-ray data analysis package CIAO v3.1.  We created exposure maps
for the images using the CIAO script {\it
merge\_all},\footnote{http://cxc.harvard.edu/ciao/download/scripts/merge\_all.tar}
and we found and measured positions, position errors, and 0.3-10 keV
fluxes for the sources in the image using the CIAO task {\it
wavdetect}.\footnote{http://cxc.harvard.edu/ciao3.0/download/doc/detect\_html\_manual/Manual.html}
The positions and errors from the first two detections of the new
X-ray transient found in the 26-Nov-2003 ACIS-I (ObsID 4679)
observation are given in Table~\ref{xobs}.  Each data set detected
sources down to (0.3-10 keV) fluxes of $\sim$8$\times$10$^{-6}$
photons cm$^{-2}$ s$^{-1}$.

We cross-correlated the X-ray source positions of all 3 observations
against all previously published X-ray catalogs and the {\it
Simbad}\footnote{http://simbad.u-strasbg.fr/} database to look for any
new, bright X-ray source likely to be an XRN.  We found several new
X-ray sources in the data.  Herein we focus on one bright new source
in particular at R.A.= 00:42:56.038, Dec.=41:12:18.50, which we name
CXOM31~J004256.0+421218, following the naming convention described in
\citet{kong2002}.  We also give the source a short name, r2-70, which
is derived from the position using the description given in
\cite{williams2004}, Table 1.  The source is 2.1$'$ east and 3.8$'$
south of the nucleus.

We determined the position of r2-70, discovered in the data set from
27-Nov-2003 (ObsID 4679), by aligning the observation with the
coordinate system of the Local Group Survey (LGS;
\citealp{massey2001}). The LGS images have an assigned J2000 (FK5)
world coordinate system accurate to $\sim$0.25$''$, and they provided
the standard coordinate system to which we aligned all of our data for
this project.  We aligned the positions of 6 X-ray sources with known
globular cluster counterparts to the positions of the centers of their
host globular clusters in the images of the LGS using the
IRAF\footnote{IRAF is distributed by the National Optical Astronomy
Observatory, which is operated by the Association of Universities for
Research in Astronomy, Inc., under cooperative agreement with the
National Science Foundation.} tasks {\it imcentroid} and {\it ccmap}.

We repeated this alignment process for the second detection of the
source on 27-Dec-2003 (ObsID 4680; see Figure~\ref{ims}) and for the
third detection of the source on 31-Jan-2004 (ObsID 4681).  Using the
independent position measurements for these detections, which were the
three detections with the highest number of counts, allowed checks for
consistency as well as the ability to reduce the final errors in the
position of the X-ray source.  The alignment errors between the {\it
Chandra} and LGS are shown for each observation in column
$\sigma_{AL}$ in Table~\ref{xpos}; random position errors for the
source, as measured by {\it wavdetect} are given in column
$\sigma_{pos}$.  Alignments allowed for adjustments in pixel scale as
well as rotation and shifts in $X$ and $Y$.

We extracted the X-ray spectrum of r2-70 from all 4 detections using
the CIAO task {\it
psextract}\footnote{http://cxc.harvard.edu/ciao/ahelp/psextract.html}.
We then fit these spectra independently, binning so that each energy
bin contained $\sim$10 counts.  These binning factors allowed the use
of standard $\chi^2$ statistics when fitting the spectra.

We fit the spectra with an absorbed power-law model and an absorbed
disk blackbody model using the CIAO 3.1/Sherpa fitting package
\citep{freeman2001}.  The best fitting model parameters, and the
associated fitting statistics, are provided in Tables~\ref{pl}
and \ref{mcd}.  Results are discussed in \S~\ref{results}.

\section{Optical Data}

{\it HST} ACS data were obtained at UT 21:37 on 25-Dec-2003 and one at
UT 04:19 on 02-Oct-2004.  Each of these were pointed at
R.A.=00:42:56.03, Dec.=41:12:19.0.  The observations had orientations
of 60.10 deg and 169.02 deg respectively.  Both observations were
taken using the standard ACS box 4-point dither pattern to allow the
final data to be drizzled to recover the highest possible spatial
resolution.  All exposures were taken through the F435W filter.  The
total exposure times were 2200 seconds for each data set.

\begin{figure}
\centerline{\psfig{file=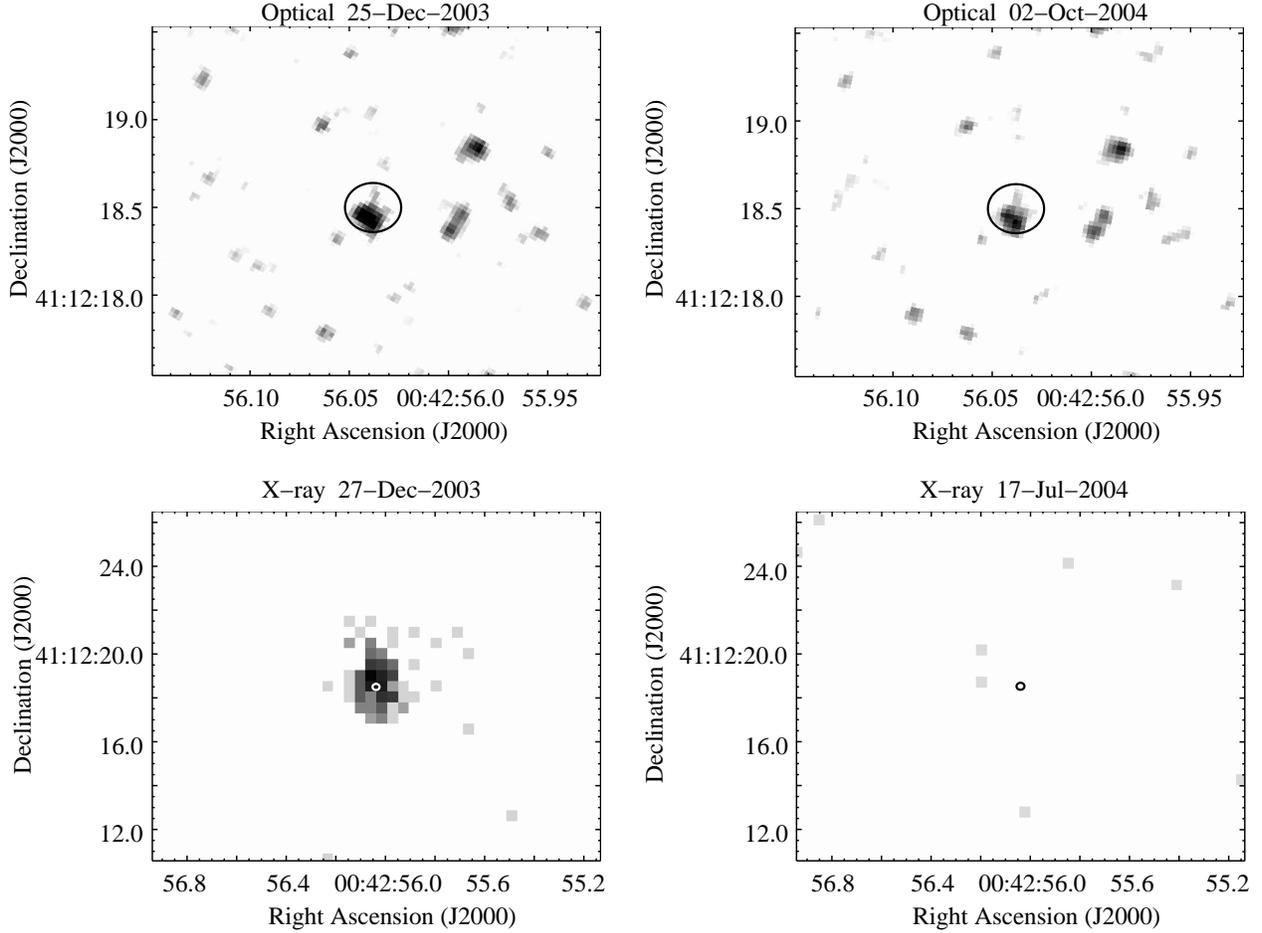,width=6.5in,angle=0}}
\caption{{\it Top left panel:} The combined 2$\sigma$ ($\pm$0.16$''$
in R.A. and $\pm$0.14$''$ in decl.) X-ray position errors for r2-70
are shown with a black ellipse on the {\it HST} image from 25-Dec-2003.
The optical counterpart candidate is the northeast portion of the
bright blend in the south part of the error ellipse.  {\it Top right
panel:} The same error ellipse is shown on the {\it HST} image from
02-Oct-2004.  The blend is now well-resolved, and the northeast
component has faded.  {\it Lower left panel:} The ACIS-I image of
r2-70 from 27-Dec-2003.  The white ellipse marks the best position for
the X-ray source in this detection.  {\it Lower right panel:} The same
error ellipse is shown on the ACIS-I image from 17-July-2004.  Source
r2-70 is not detected.}
\label{ims}
\end{figure}

We aligned and drizzled each set of 4 images into high-resolution
(0.025$''$ pixel$^{-1}$) images using the PyRAF\footnote{PyRAF is a
  product of the Space Telescope Science Institute, which is operated
  by AURA for NASA.} task {\it multidrizzle},\footnote{multidrizzle is
  a product of the Space Telescope Science Institute, which is
  operated by AURA for
  NASA. http://stsdas.stsci.edu/pydrizzle/multidrizzle} which has been
optimized to process ACS imaging data.  The task removes the cosmic
ray events and geometric distortions, and it drizzles the dithered
frames together into one final photometric image.  Sections of the
images, centered on the position of the transient X-ray source, are
shown in Figure~\ref{ims}.

The drizzled ACS images were aligned with the LGS coordinate system
with {\it ccmap} using stars common to both images.  The alignment
errors were $\sim$0.04$''$, indicating that the ACS images were
accurately aligned to the LGS system.  With the ACIS-I and ACS images
aligned to the same coordinate system, we were able to compare the
coordinates of the ACS and ACIS-I sources reliably.  The X-ray
position error ellipse for r2-70 is shown on the aligned $HST$ images
in Figure~\ref{ims}.  The images reveal a fading optical source at the
same position as the transient X-ray source, indicating that we
detected the optical counterpart of the X-ray event.  However, the
counterpart has a bright neighbor which is apparent in the second
$HST$ image.  This neighbor led to some complications in measuring the
photometry of the counterpart.

We processed the relevant sections of the final images with DAOPHOT-II
and ALLSTAR \citep{stetson} to find and measure the count rates of the
optical counterpart.  We converted the count rates measured on our
images to VEGA magnitudes using the conversion techniques provided in
the ACS Data
Handbook\footnote{http://www.stsci.edu/hst/acs/documents/handbooks/DataHandbookv2/ACS\_longdhbcover.html}.

\section{Results}\label{results}

\subsection{X-ray}

The brightest flux at which the source was observed was
1.4$\times$10$^{-4}$ photon cm$^{-2}$ s$^{-1}$.  Since this source is in
the region surveyed by \citet{kong2002} to a detection limit of
$\sim$8$\times$10$^{-7}$ photon cm$^{-2}$ s$^{-1}$, the source
demonstrates changes in flux of more than a factor of 100, indicating
that the source was an X-ray Nova.

The results of the spectral fits to the 4 detections of r2-70 are
given in Tables~\ref{pl} and \ref{mcd}.  All of these fits agree that
r2-70 had a soft spectrum; however, the measured absorption is much
higher for the power-law fits.  In fact, the best fits using the disk
blackbody model have no absorption, which is not possible considering
the known Galactic foreground absorption toward M31
($\sim$6$\times$10$^{20}$ cm$^{-2}$).  This result, along with the
slightly lower quality of the disk blackbody fits, favors the
power-law as the correct spectral model, and an absorption-corrected
X-ray luminosity of 6$\times$10$^{37}$ erg s$^{-1}$.  As this
luminosity is below the Eddington luminosity of a 1.4 M$_{\odot}$
neutron star, the luminosity does not discriminate between an
accreting neutron star and an accreting black hole.

\citet{tanaka1996} discuss the differences between the X-ray spectra
of accreting neutron stars and those of accreting black holes. Both
types of accreting objects have two-component spectra.  Neutron stars
have a blackbody component consistent with the neutron star surface
and a multi-temperature disk blackbody component.  The disk component
typically has a maximum temperature of kT$\sim$1.5 keV at 10$^{38}$
erg s$^{-1}$ which decreases as the luminosity goes down.  Black holes
have a power-law component and a multi-temperature disk blackbody
component with a peak temperature of kT$<$1.2 keV.  The key difference
is the presence of a blackbody component consistent with a neutron
star surface.  If this component is not present, then the source is
more likely a black hole.  We fitted the X-ray spectrum with the most
counts (ObsID 4681) with both of these two-component models and both
had very good ($\chi^2/dof=1.0$) fits.  We therefore cannot rule out
either possibility based on the coarse spectra from the discovery
observations.

\begin{figure}
\centerline{\psfig{file=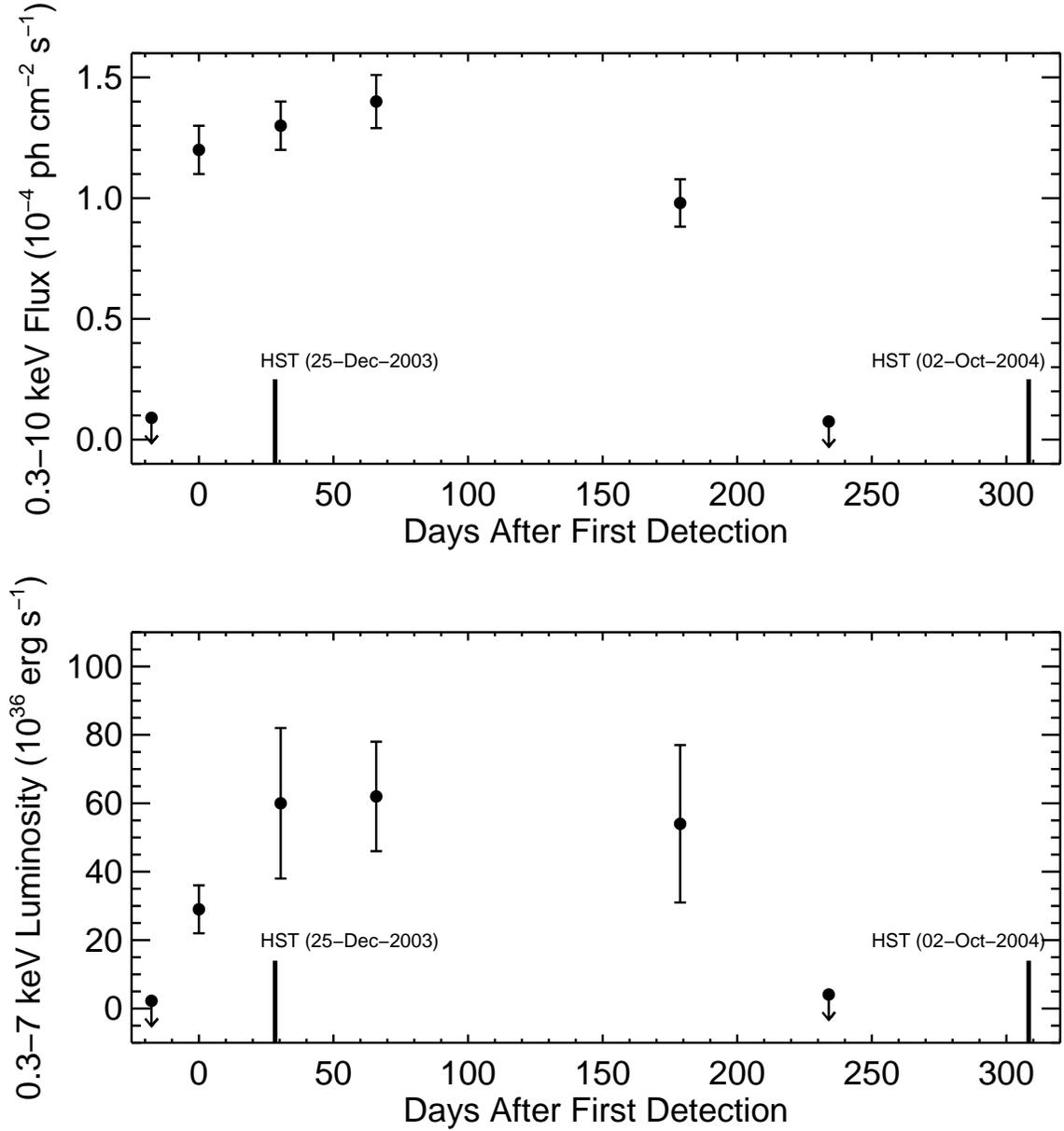,width=6.5in,angle=0}}
\caption{{\it Top panel:} The 0.3-10 keV X-ray flux lightcurve for
r2-70. Error bars are 1$\sigma$.  Upper-limits are 3$\sigma$. {\it
Bottom panel:} The 0.3-7 keV X-ray luminosity lightcurve for
r2-70. As in the top panel, labeled long vertical ticks show the
timing of our $HST$ observations.}
\label{lc}
\end{figure}

The X-ray flux and X-ray luminosity lightcurves are shown in
Figure~\ref{lc}.  The 3$\sigma$ upper-limit shown for the 17-Jul-2004
observation is representative of the {\it Chandra} observations
through October of 2004; none of which detected r2-70.  Because the
X-ray flux began to decay sometime during the 4 months when we have no
observations, we were unable to constrain the shape of the decay curve
of the X-ray transient.  It is clear from the many detections that
this transient source was active for at least 6 months.  The source
also appears to have decayed rapidly once the decay started, fading
from 10$^{-4}$ photon cm$^{-2}$ s$^{-1}$ to a 3$\sigma$ upper limit of
$<$7.5$\times$10$^{-6}$ photon cm$^{-2}$ s$^{-1}$ in 55 days.  If an
exponential decay is assumed, this decay by at least a factor of 13.3
indicates an $e$-folding decay time of $\lap$21 days, typical for
Galactic X-ray novae \citep{chen1997}.

\subsection{Optical}

Inside of the 2 $\sigma$ {\it Chandra} error ellipse on the first ACS
image, DAOPHOT found 3 optical point sources.  The brightest two of
these are in the southern portion of the error ellipse.  These sources
are within 0.1$''$ of each other.  The significant fading of the
northeastern source between $HST$ observations, as seen in
Figure~\ref{ims}, distinguished it as an optical counterpart candidate
for the transient X-ray event.  Another source lies to the 0.15$''$ to
the north of the blended sources.  It had magnitudes of
$B=25.95\pm0.09$ and $B=25.97\pm0.05$ in the first and second ACS
images respectively.  Since this source did not fade between
observations, we did not consider it further as a counterpart
candidate, leaving the northeastern member of the blended pair as our
only counterpart candidate.

Distinguishing the photometry of the pair of bright optical sources,
separated by only 0.10$''$, was difficult in the 25-Dec-2003 data set
because the brighter northeastern source overpowered the fainter
southwestern source.  However, in the 02-Oct-2004 data set, the
northeastern source had faded sufficiently that ALLSTAR was able to
cleanly measure the photometry for both sources individually, finding
the northeastern source to be $B=24.92\pm0.06$ and the southwestern
source to be $B=24.90\pm0.05$.  Therefore, assuming typical extinction
to M31 ($A_B$=0.4) and a distance modulus of 24.47
\citep{williams2003}, the brightest stars consistent with the position
of r2-70 have M$_B\sim 0$, fainter than the O and B stars found in
HMXBs.  Since most Galactic SXTs in LMXBs contain black holes, the
soft X-ray spectrum and lack of high-mass stars at the position of
r2-70 suggest that it is a black hole binary.

We measured the total count rate of the blended sources in the first
ACS image, using an aperture of radius 0.14$''$ centered on the blend
and subtracting the background level sampled in an annulus from
0.30$''$ to 0.55$''$.  By subtracting the contribution of the light
expected from the $B=24.90$ southwestern neighbor, we obtained the
count rate of the northeastern source alone.  This technique yielded
$B=24.52\pm0.07$ for the northeastern source during the first $HST$
observation.  Assuming a distance modulus to M31 of 24.47
\citep{williams2003} and extinction consistent with $N_H$, the
absolute magnitude of the optical counterpart was M$_B$=-1.3$\pm$0.5.

For consistency, we applied the same aperture photometry technique to
the second observation of the pair of optical sources.  The technique
yielded $B=24.95\pm0.08$ for the northeastern source during the second
$HST$ observation.  This measurement is equivalent at the 1$\sigma$
level with the ALLSTAR measurement of the northeastern source in the
second image, suggesting that this technique was successful at
separating the photometry of the blended sources.

The aperture photometry reveals an optical source inside of the X-ray
position error ellipse that faded by 30\% from the time the X-ray
source was active to the time that the X-ray source was quiescent.
This optical variability in concert with the X-ray source demonstrated
by the brightest of the four optical sources detected in the error
ellipse clearly indicates that this fading optical source was the
counterpart to the transient X-ray event.

The bright optical luminosity ($B$=25) apparent during X-ray
quiescence may have several explanations.  For example, r2-70 could be
an intermediate mass X-ray binary, similar to V4641~Sgr, whose binary
companion is a late B or early A star \citep{chaty2003} and whose
period is 2.8 days \citep{orosz2001}.  In this case, the optical light
from the binary companion will contaminate the optical light from the
accretion disk even during the X-ray outburst, and the optical
brightening during outburst is typically only 1-2 mag (see references
in \citealp{orosz2001}).  Another example of a black hole binary that
is optically bright during quiescence is 4U~1543-47.  While this
source has had peak X-ray luminosities of $>10^{39}$ erg s$^{-1}$
\citep{park2004} and quiescent X-ray luminosities of $<10^{31.5}$ erg
s$^{-1}$ \citep{garcia2001xn}, its optical brightness only changes by
$\sim$1.8 mag \citep{orosz1998,vanparadijs1995}.  This system also has
an early A star secondary and a period of $>$1 day (27 hr;
\citealp{orosz1998}).  With its long predicted period of $\sim$8 days
(see \S 4.3), bright quiescent optical luminosity, and modest
brightening during outburst, r2-70 bears a strong resemblance to these
Galactic X-ray binaries.

On the other hand, it is also possible that both of the $B$=25 stars
visible during X-ray quiescence are chance superpositions.  In this
case, the optical counterpart of the LMXB was only seen during the
outburst but was not resolved because of its proximity to this pair of
stars.  Either of these possibilities would cause the optical
luminosity of the accretion disk during outburst to be somewhat
overestimated.

\subsection{Orbital Period Prediction}

We applied the relation of \citet{vanparadijs1994} to predict the
orbital period of the LMXB that produced the X-ray transient we have
detected as r2-70.  This empirical relation seen in Galactic LMXBs
shows that the X-ray/Optical luminosity ratio of an LMXB is correlated
with the orbital period of the system.  The relation appears to hold
even for more recently discovered systems \citep{williams2005a}.  With
our optical and X-ray data of r2-70, we are able to measure the X-ray
and optical luminosities.  Assuming that r2-70 is an LMXB, as is
suggested by its location in the M31 bulge, which contains very few,
if any, young, high-mass stars (e.g. \citealp{stephens2003}), and
assuming that LMXBs in M31 follow the same relation between their
photometric properties and their orbital periods as those in the
Galaxy, our luminosity measurements for r2-70 put a constraint on the
orbital period of the system.

To determine the $V$-band luminosity of r2-70 during the outburst, we
first took our apparent $B$ magnitude of $24.52\pm0.10$ for r2-70
during the first ACS observation.  We then determined the extinction
toward r2-70 using the fits to the X-ray spectra.  Taking the weighted
mean of the 4 independent measurements of the column density given in
Table~\ref{pl}, we obtained $N_H=(1.8\pm0.5)\times 10^{21}$ cm$^{-2}$.
Using the relation of \citet{predehl1995} and the standard
interstellar extinction law, this absorption translates to $A_B =
1.3\pm0.4$.  Applying an intrinsic $B-V$ color of $-0.09\pm0.14$
determined from the Galactic LMXB catalog of \citet{liu2001}, and a
distance of 780 kpc, our measurement of M$_V = -1.2\pm0.4$.

The X-ray luminosity was determined using the power-law spectral fits
to the ACIS data (see Table~\ref{pl}).  The absorption-corrected
0.3-7 keV X-ray luminosity of r2-70 on 27-Dec-2003, as measured from
the {\it Chandra} data, was of $(6\pm2)\times 10^{37}$ erg s$^{-1}$.
Assuming this was the X-ray luminosity during the 25-Dec-2003 $HST$
observation, this X-ray luminosity and the optical luminosity of M$_V
= -1.2\pm0.4$, measured from the 25-Dec-2003 $HST$ data, can be
applied to the relation of \citet{vanparadijs1994}.  This calculation
yields a predicted orbital period of 8$^{+12}_{-5}$ days for this
system.  

We note that if the lower extinction (2.5$\times 10^{20}$ cm$^{-2}$)
and X-ray luminosity ((2.5$\pm$1.3)$\times 10^{37}$ erg s$^{-1}$) from
the disk blackbody spectral fits are applied, M$_V$ = 0.0$\pm$0.2 on
25-Dec-2003 and the orbital period prediction changes to 3$^{+3}_{-2}$
days.  We also note that if the optical luminosity of the accretion
disk was actually lower than the measured value because the
counterpart is an A star or because the counterpart was poorly
resolved from its neighbors, the range of the predicted orbital period
would decrease.

Comparisons with the other two optical detections of X-ray transients
in M31 show that the properties of the optical counterpart for r2-70
were between those of the other transient counterparts.  The
counterpart for r2-70 was fainter than that seen for the X-ray
transient r2-67 (M$_V$=-2.4$\pm$0.8; \citealp{williams2004}) and
brighter than that seen for the counterpart candidate for the X-ray
transient s1-86 (M$_V$=-0.25$\pm$0.27; \citep{williams2005a}).  The
predicted orbital period also falls between the values calculated for
r2-67 (23$^{+54}_{-16}$ days; \citealp{williams2004}) and s1-86
(1.0$^{+2.9}_{-0.6}$ days; \citealp{williams2005a}).

\section{Conclusions}

We have discovered a new transient X-ray source in the M31 bulge that
appeared in November of 2003, which we have named r2-70.  This source
attained an X-ray flux more than a factor of 100 greater than the
upper-limits of previous surveys in which it did not appear.  The
transient event kept a high X-ray flux through the first half of 2004.
When the source decayed, it did so with an $e$-folding decay time of
$\lap$1 month.  The event had a soft spectrum best fit by a power-law
with index $\sim$3, and it had an absorption-corrected 0.3-7 keV
luminosity $\sim$6$\times$10$^{37}$ erg s$^{-1}$.

Follow-up $HST$/ACS F435W ($B$-band equivalent) imaging revealed a
fading optical source within the tightly constrained error ellipse of
the location of the transient X-ray event, showing that this optical
source is the optical counterpart of the X-ray transient.  The source
decayed from $B=24.52\pm0.07$ to $B=24.95\pm0.08$ between epochs.
Assuming that the transient event occurred in an LMXB, we can apply
our X-ray and optical luminosity measurements to the empirical
relation of \citet{vanparadijs1994} to predict that the system has an
orbital period of 8$^{+12}_{-5}$ days.

Finally, the lack of high-mass stars at the position of r2-70 suggests
that it is an LMXB.  Although the X-ray luminosity and spectrum do not
exclude either an accreting neutron star or black hole, many Galactic
LMXBs that exhibit such transient events and have such soft X-ray
spectra contain stellar mass black holes, making r2-70 a good black
hole candidate in M31.

Support for this work was provided by NASA through grant number
GO-9087 from the Space Telescope Science Institute and through grant
number GO-3103X from the {\it Chandra} X-Ray Center.  MRG acknowledges
support from NASA LTSA grant NAG5-10889.

%\bibliography{apjmnemonic,references}
%\bibliographystyle{apj}

\clearpage

\begin{deluxetable}{ccccccccccc}
%\tablewidth{in}
\tablecaption{{\it Chandra} ACIS-I observations}
\tableheadfrac{0.01}
\tablehead{
\colhead{{ObsID}} &
\colhead{{Date}} &
\colhead{{R.A. (J2000)}} &
\colhead{{Dec. (J2000)}} &
\colhead{{Roll (deg.)}} &
\colhead{{Exp. (ks)}}
}
\startdata
4678 & 09-Nov-2003 & 00 42 44.4 & 41 16 08.3 & 239.53 & 3.9\\
4679 & 26-Nov-2003 & 00 42 44.4 & 41 16 08.3 & 261.38 & 3.8\\
4680 & 27-Dec-2003 & 00 42 44.4 & 41 16 08.3 & 285.12 & 4.2\\
4681 & 31-Jan-2004 & 00 42 44.4 & 41 16 08.3 & 305.55 & 4.1\\
4682 & 23-May-2004 & 00 42 44.4 & 41 16 08.3 & 79.99 & 3.9\\
4719 & 17-Jul-2004 & 00 42 44.3 & 41 16 08.4 & 116.83 & 4.1\\
\enddata
\label{xobs}
\end{deluxetable}
\begin{deluxetable}{ccccccccc}
\tablewidth{7in}
\tablecaption{Position Measurements and Errors of XRN r2-70}
\tableheadfrac{0.01}
\tablehead{
\colhead{{ID}} &
\colhead{{R.A. (J2000)}} &
\colhead{{$\sigma_{pos}$\tablenotemark{a}}} &
\colhead{{$\sigma_{AL}$\tablenotemark{b}}} &
\colhead{{$\sigma_{tot}\tablenotemark{c}$}} &
\colhead{{Dec. (J2000)}} &
\colhead{{$\sigma_{pos}$}} &
\colhead{{$\sigma_{AL}$}} &
\colhead{{$\sigma_{tot}$}}
}
\tablenotetext{a}{Random position errors were measured using the CIAO task {\it wavdetect}.}
\tablenotetext{b}{Errors in the alignment between the X-ray image and the LGS coordinate system were measured using the IRAF task {\it ccmap}.}
\tablenotetext{c}{Total position errors were calculated by adding the position and alignment errors in quadrature.}
\tablenotetext{d}{The mean position and errors were calculated using standard statistics for combining multiple measurements \citep{bevington}.}
\startdata
4679 & 00 42 56.025 & 0.09$''$ & 0.08$''$ & 0.12$''$ & 41 12 18.74 & 0.08$''$ & 0.11$''$ & 0.14$''$\\
4680 & 00 42 56.032 & 0.06$''$ & 0.19$''$ & 0.20$''$ & 41 12 18.61 & 0.06$''$ & 0.14$''$ & 0.15$''$\\
4681 & 00 42 56.056 & 0.06$''$ & 0.12$''$ & 0.13$''$ & 41 12 18.30 & 0.06$''$ & 0.09$''$ & 0.11$''$\\
\hline
Mean\tablenotemark{d} & 00 42 56.038 & \nodata & \nodata & 0.08$''$ & 41 12 18.50 & \nodata & \nodata & 0.07$''$\\
\enddata
\label{xpos}
\end{deluxetable}

\begin{deluxetable}{cccccccccc}
\tablewidth{7in}
\tablecaption{Power-law spectral fits}
\tableheadfrac{0.01}
\tablehead{
\colhead{{Date}} &
\colhead{{Cts\tablenotemark{a}}} &
\colhead{{Flux\tablenotemark{b}}} &
\colhead{{Slope\tablenotemark{c}}} &
\colhead{${\rm N_H}\tablenotemark{d}$} &
\colhead{{$\chi^2/\nu$}} &
\colhead{{Q\tablenotemark{e}}} &
\colhead{{{HR-1\tablenotemark{f}}}} &
\colhead{{{HR-2\tablenotemark{g}}}} &
\colhead{{$\rm L_X$\tablenotemark{h}}}
}
\tablenotetext{a}{The background-subtracted number of counts in the detection.}
\tablenotetext{b}{The exposure corrected 0.3-10 keV flux in units of 10$^{-4}$
photon cm$^{-2}$ s$^{-1}$.}
\tablenotetext{c}{Slope of the best-fitting absorbed power law model.}
\tablenotetext{d}{The absorption column in units of 10$^{21}$ cm$^{-2}$.}
\tablenotetext{e}{The probability that this fit is representative of
the true spectrum, determined from $\chi^2/dof$.}
\tablenotetext{f}{Hardness ratio calculated by taking the ratio of
M-S/M+S, where S is the number of counts from 0.3-1 keV and M is the
number of counts from 1-2 keV.}
\tablenotetext{g}{Hardness ratio
calculated by taking the ratio of H-S/H+S, where S is the number of
counts from 0.3-1 keV and H is the number of counts from 2-7 keV.}
\tablenotetext{h}{The absorption-corrected luminosity of the source in
units of 10$^{36}$ \ergs (0.3-7 keV).}
\startdata
26-Nov-2003 & 134 & 1.2$\pm$0.1 & 1.7$\pm$0.3 & 0.6$\pm$0.9 & 7.08/10 & 0.72 & 0.38$\pm$0.11 & 0.24$\pm$0.12 & 29$\pm$7 \\
27-Dec-2003 & 145 & 1.3$\pm$0.1 & 3.1$\pm$0.5 & 2.6$\pm$1.2 & 17.84/11 & 0.09 & 0.33$\pm$0.09 & -0.19$\pm$0.12 & 60$\pm$22\\
31-Jan-2004 & 161 & 1.4$\pm$0.1 & 3.1$\pm$0.3 & 2.1$\pm$0.9 & 10.46/12 & 0.58 &  0.24$\pm$0.08 & -0.47$\pm$0.13 & 62$\pm$16\\
23-May-2004 & 109 & 1.0$\pm$0.1 & 3.3$\pm$0.6 & 2.6$\pm$1.4 & 7.93/7 & 0.34 &  0.31$\pm$0.11 & -0.42$\pm$0.16 & 54$\pm$23\\
\enddata
\label{pl}
\end{deluxetable}

\begin{deluxetable}{ccccccc}
\tablewidth{6.5in} \tablecaption{Disk Blackbody Spectral Fits to ACIS-I Detections of r2-70}
\tableheadfrac{0.01}
\tablehead{
\colhead{{Date}} &
\colhead{{T$_{in}$\tablenotemark{a}}} &
\colhead{{R$_{in}cos^{1/2}i$ \tablenotemark{b}}} &
\colhead{{${\rm N_H}$\tablenotemark{c}}} &
\colhead{{$\chi^2/\nu$}} &
\colhead{{Q\tablenotemark{d}}} &
\colhead{{$\rm L_X$\tablenotemark{e}}} }
\tablenotetext{a}{The temperature of the inner disk in keV.}
\tablenotetext{b}{The radius of the inner disk in km, assuming the distance to M31 is 780 kpc and the inclination ($i$) of the binary is 0 degrees.}
\tablenotetext{c}{The absorption column in units of 10$^{21}$ cm$^{-2}$.}
\tablenotetext{d}{The probability that this fit is representative of
the true spectrum, determined from $\chi^2/dof$.}
\tablenotetext{e}{The absorption-corrected luminosity of the source in
units of 10$^{36}$ \ergs (0.3-7 keV).}
\tablenotetext{f}{The absorption was fixed to fit the spectrum from 23-May-2004, as $N_H$ was unconstrained if left as a free parameter.}
\startdata
26-Nov-2003 & 1.5$\pm$0.2 & 5$^{+1}_{-2}$ & $\leq$0.1 & 8.65/10 & 0.57 & 25$\pm$13\\
27-Dec-2003 & 0.5$\pm$0.1 & 30$^{+13}_{-30}$ & 0.3$\pm$0.8 & 23.26/11 & 0.02 & 20$\pm$20\\
31-Jan-2004 & 0.58$\pm$0.07 & 28$^{+7}_{-9}$ & $\leq$0.1 & 12.82/12 & 0.38 & 20$\pm$10\\
23-May-2004 & 0.50$\pm$0.07 & 31$^{+9}_{-12}$ & 0.25\tablenotemark{f} & 10.30/8 & 0.24 & 14$\pm$9\\
\enddata
\label{mcd}
\end{deluxetable}

\end{document}